# Cystic Lung Phantom to Validate Clinical CT Protocols

Shefra Shah [1], Farah Hussaini [1], Dumitru Mazilu [1], Eric Bennett [1], and Han Wen [1,*]

1 Laboratory of Imaging Physics, Biochemistry and Biophysics Center, National Heart, Lung and Blood Institute, National Institutes of Health, Bethesda, MD 20892, USA; shefra.shah@nih.gov (S.S.) farah.hussaini@nih.gov (F.H.); mazilud@nhlbi.nih.gov (D.M.); bennette@nhlbi.nih.gov (E.B.)
* Correspondence: wenh@nhlbi.nih.gov

**Abstract:** In CT-based evaluation of the extent of cystic changes in the lungs of patients with cystic lung diseases, such as Lymphangioleiomyomatosis (LAM), there is a lack of a lung phantom containing air-filled cavities that mimic pulmonary cysts to calibrate the measurement of cystic volumes from CT scans. Here we describe a simple, easy-to-replicate cystic lung phantom consisting of basic structures of a trachea and two lung compartments. The lung compartments contain air cavities of varying sizes to mimic cystic lesions. The lung volumes are equal to those of typical adults. The lung compartments are made of a foam material recommended by NIST to simulate the radiodensity of human lung parenchyma. In tests performed on a clinical scanner using two types of lung analysis software, the various structures in the lung phantom were correctly recognized by the software. The resulting cystic volume measurements revealed the relationship between the size of the cysts and the accuracy of the measurement. A significant finding is that the volumes of individual cysts were underestimated for small cysts. The error increased with decreasing cyst sizes. Such underestimation has not been mentioned previously and deserves the attention of clinicians using CT scans to assess the cyst burden in the lungs, particularly in patients presenting with numerous small pulmonary cysts.

## 1. Introduction

Lymphangioleiomyomatosis (LAM), a rare, progressive lung disease predominantly found in women of childbearing age, is caused by the proliferation of abnormal smooth muscle-like cells, leading to wall lung tissue thickening, the development of air-filled lung cysts, and an overall decline in pulmonary function LAM affects approximately three to eight million women worldwide with significant variation, yet, many patients with LAM remain undiagnosed [1-4]. LAM exists in two forms: sporadic LAM and tuberous sclerosis complex LAM (TSC-LAM). Sporadic LAM is associated with lymphatic (e.g. chylous pleural effusions, thoracic duct dilations) manifestations in a non-hereditary and randomized fashion [3-5]. TSC-LAM results from mutations in genes *TSC1* and *TSC2*, where it commonly manifests in the brain (e.g. seizures), renal (e.g. angiomyolipomas), and skin (e.g. lesions) divisions [3-5]. Computed tomography (CT) measurements of the cystic lung changes over time are a standard measure to assess the progression of the disease and aid in treatment decisions.

Diagnosing LAM can be challenging due to its rarity. Fortunately, CT scans aid in the diagnosis of LAM, providing a high-resolution view of the cystic lung structure. Moreover, CT-derived quantitative assessments of pulmonary cysts, resulting in a percentage of the total lung volume occupied by cysts (cyst score), serve as a useful metric in monitoring cystic changes.

The current gold-standard cyst segmentation method is calculated by an FDA-approved semi-automatic software such as the Lung Volume Analysis (Canon Medical Systems USA, Inc. Tustin, CA, USA). In addition to this method, fully automatic cyst segmentation method can be used to calculate cyst measurements for each CT lung scan [6]. To assess the accuracy of cyst volume measurements from CT scans, verification is required. Verification tools, such as a lung phantom, can be used to establish imaging systems. However, current generalized lung phantoms typically implement solid lesions rather than air-filled cysts, which would not meet the need of CT calibrations for diseases such as LAM. In the meantime, there has been a substantial effort from the National Institute of Standards and Technology (NIST) to identify commercially available polymer foam as low-cost material that mimics the radiodensity of lung parenchyma in CT scans [7].

The purpose of this study was to engineer a cystic lung phantom, using the NIST lung phantom material, to verify the CT pipeline of cyst measurements. Calibrating these measurements will allow for greater confidence in current cyst scoring methods.

## 2. Materials and Methods

A commercially available polymer foam has been used in a previous study on identifying a low-cost density reference phantom for CT lung imaging by the National Institute of Standards and Technology (NIST) [7]. Levine *et al.* have demonstrated that the polymer foam serves as a density range for lung parenchyma and can be implemented in cystic lung phantom models.

The average lung capacity for a pair of lungs in a male and female is about 6 liters and 4 liters, respectively [5,8,9]. Although almost all LAM patients are female, there are occasional male patients who exhibit the disease or another form of cystic lung disease. We constructed our cystic lung phantom to account for an average between these particular lung capacities, totaling a volume of approximately 5 liters in both lung compartments.

In our study, we used a polymer foam suggested by NIST [7] to fill the lung volumes in our cystic lung phantom model. Two rectangular foam blocks, each with the dimensions of 204mm x 98mm x 125mm, were used to model the right and left lung compartments. In the interior of each lung compartment (Figure 1), 12 cylindrical cavities were cut using a CNC system. Cavities are arranged in four rows and three columns. The cavities include varying sizes: three with a diameter and depth of 20 mm, three with a diameter and depth of 15 mm, three with a diameter and depth of 10mm, and three with a diameter and depth of 5 mm. These cavities model cysts of varying sizes in the lungs. Figure 2 shows a 3D rendition of the cavities in the lung compartments.

Cast acrylic sheets were used for the exterior walls of the cystic lung phantom (Figure 2). Figure 3 shows a coronal cut-away view of the phantom. The outer dimensions of each lung box were 254.8 mm x 259.5 mm x 150.4 mm. The two compartments are separated by an acrylic wall. To mimic the trachea, a rectangular channel of 12.7 mm width was cut in the middle of the septal wall between the left and right lung compartments, extending to half the length of the phantom. The distal end of the channel opens to both lung compartments.

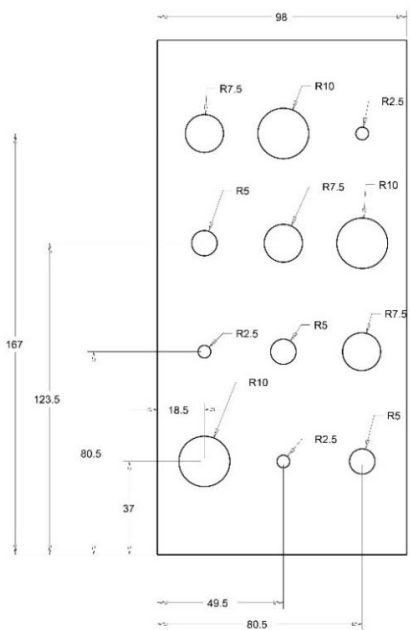

**Figure 1.** Design of interior cavities in a polymer foam block that fills a lung compartment. These cavities mimic pulmonary cysts. There are a total of 12 cavities in 4 groups of different sizes.

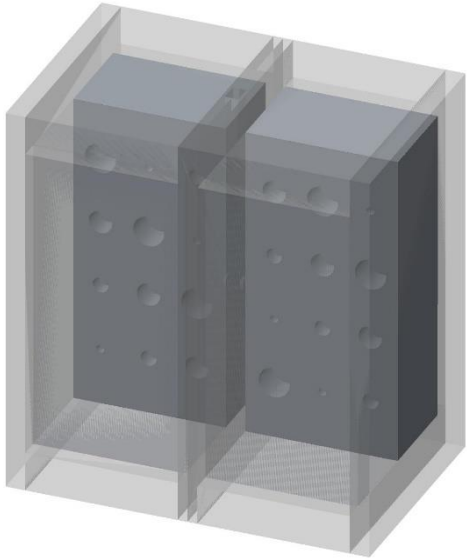

**Figure 2.** Three-dimensional representation of the lung phantom. The outer walls of cast acrylic are shown in light gray. The polymer foam with cavities within the lung compartments is shown in dark gray, in a cut-away view.

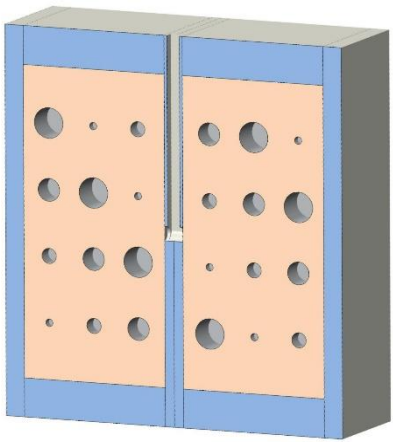

**Figure 3.** A coronal cut-away view of the cystic lung phantom. The orange blocks with 24 cavities represent the lung compartments, while the blue surrounding represents the cast acrylic material. Additionally, the channel in the septal wall mimicking the trachea is at clear view, in between the two orange compartments and ends at about ½ of the phantom length.

The cystic lung phantom was scanned on a clinical CT scanner (Canon Aquillion One Genesis, Canon Medical Systems USA, Inc. Tustin, CA, USA) using a low-dose chest scan protocol for LAM patients. The scan protocol was a helical scan at 120kVp/R700 mA, with a total scan time of 4.37 seconds. The dose-length product of the scan was nominally 90.90 mGy · cm, and the local intensity of radiation was 3.3 mGy measured by the CTDIvol index [3,6,10].

The chest CT scan was reconstructed to a nominal 360 mm field of view, 512 by 512 matrix size, and 1-mm slice thickness and interval. For cyst segmentation and measurements with the commercial semi-automatic software, the operator has the option of manually seeding the airways, the left lung and right lung compartments. The operator also adjusts or confirms a threshold of CT signal level in Hounsfield units for the identification of air-filled cysts. This was set at -940 HU by default. With these inputs, the software will produce a 3D segmentation of the cysts and provide measurements of the total volume of the cysts in each lung. With the fully automatic software, the program generates the cyst segmentation and volume measurements without operator input.

Using the segmentation maps, we measured the volumes of individual cysts in the phantom. These were compared with the true volumes of the cysts that were calculated according to the engineering designs. The percentage errors in the CT-derived cyst volumes were calculated as the percentage fraction of (measured volume - true volume)/true volume. This was evaluated for individual cysts. To compare the levels of error between the semi-automatic and fully automatic software, we used a student's t-test to calculate the statistical significance of the difference for each group of cysts of the same size.

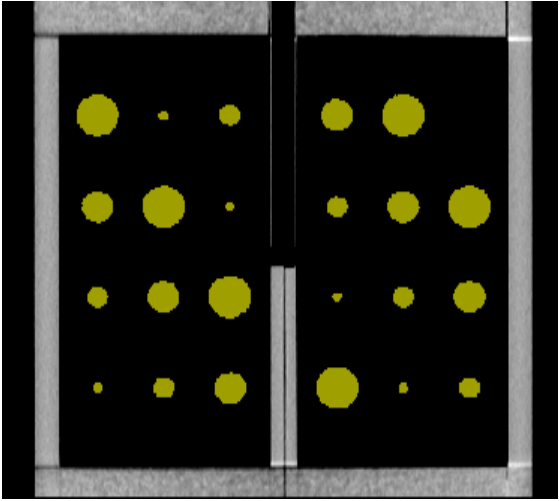

**Figure 4.** The semi-automatic segmentation software identifies the cysts of varying sizes, which are highlighted by the software in a chartreuse color. The cyst sizes from large to small are as follows: 20mm, 15mm, 10mm, and 5mm.

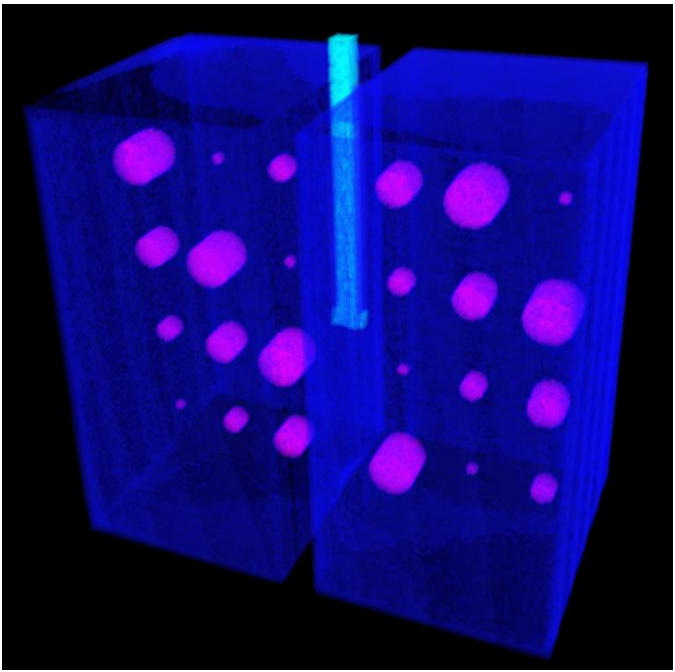

**Figure 5.** The fully automatic segmentation software displays the cysts in a magenta color and the lung compartments in the azure color. The trachea is identified by the cyan color.

## 3. Results

Both the semi-automatic and fully automatic software correctly identified the structural components of the cystic lung phantom. Figure 4 and Figure 5 show the segmentation of the various compartments by the software packages. The volume measurements accounted for the relationship between the cyst (cavity) sizes and its measurement accuracy, where six cysts of each of the same size (20mm, 15mm, 10mm, 5mm) were treated as a group in the statistical analysis. Table 1 summarizes the results of percentage error evaluation for the four groups of cysts, including the mean and standard deviation of the percentage error in each group. Also listed in Table 1 are the p-values of the statistical significance between the error levels of the semi-automatic and fully automatic measurements. Figure 6 is a visual representation of the results in a bar graph. It can be seen that as the cavity size decreases, the percentage error of the software measurements relative to the true cyst volume increases. Considering the semi-automatic software versus the automatic software, their levels of measurement error were statistically equivalent, with p-values of 0.13 or greater for all four cyst sizes (Table 1).

The two software packages also provided specific measurements of the total volume of cysts. The result was 58 $cm^3$ from the semi-automatic software, and 58.47 $cm^3$ from the automatic software. When compared to the true total volume of 58.90 $cm^3$, the measured values under-estimated by 1.5% and 0.7% respectively.

**Table 1. Percentage errors of single cyst volumes from the semi-automatic and fully automatic analysis software**

| Cyst diameters (mm) | Percentage error of single cyst volume, semi-automatic software | | Percentage error of single cyst volume, fully automatic software | |
|---|---|---|---|---|
| | Mean | Standard deviation | Mean | Standard deviation |
| 20 | -0.181% | 0.622% | -0.165% | 0.796% |
| 15 | -0.793% | 1.21% | -1.24% | 1.02% |
| 10 | -3.18% | 2.21% | -5.06% | 1.63% |
| 5 | -15.9% | 3.18% | -15.8% | 10.3% |

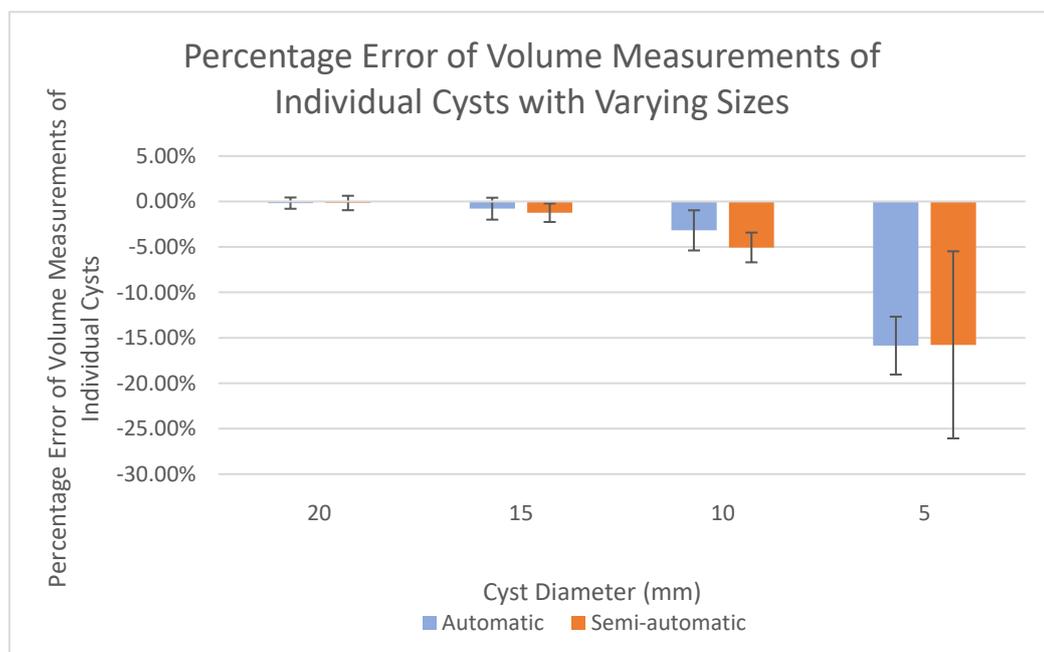

**Figure 6.** Bar graphs showing the percentage errors of volume measurements for cysts of different sizes. Results for both the automatic (blue) and semi-automatic (orange) software are plotted. The error bars are the standard deviation of the percentage error for each group of cysts of a specific size.

## 4. Discussion

In this study we constructed and tested a cystic lung phantom for the purpose of verifying the accuracy of CT-based measurements of pulmonary cysts in patients with cystic lung disease. A hallmark of cystic lung diseases, such as Lymphangioleiomyomatosis (LAM) and Birt-Hogg-Dubé (BHD) syndrome, is the presence of diffused, air-filled cysts in the lungs up to a few centimeters in size. To fulfill its purpose the requirements for the phantom are that it has the typical size of an adult's lungs, that it contains hollow cavities of varying sizes that are typical of pulmonary cysts, and that the cysts and other basic lung compartments can be recognized by software packages that are routinely used in patients.

It took several iterations to make a phantom that met the requirements, particularly one that could be correctly segmented by standard image analysis software for patients. In this process we made specific adjustments to the vertical channel modeling the trachea in the space between the two lung compartments, until the analysis software could successfully identify the lung compartments and the cysts within.

The main finding of the test was that there was a systematic under-estimation of the volume of cysts under 10 mm in diameter, and that the level of error exceeded 15% for cysts of 5 mm diameter. This was the case for both the supervised semi-automatic and unsupervised automatic measurements. It is likely that the underestimation would increase further for cysts under 5 mm size. This underestimation is hitherto unknown to our knowledge, although chest CT scans have long been the gold-standard for assessing the extent of cystic changes in the lungs [11,12]. The finding suggests that physicians should maintain caution when using CT scans to quantify the extent of cystic changes in the lungs and monitor progression of the disease, particularly in patients presenting with small pulmonary cysts. Phantom calibrations, such as this study, may also provide the possibility of a cyst-size based correction of measurements in the future.

A limitation to the phantom design is that the relatively small number of cysts in the lung compartments, with a total of 24 cysts divided into 4 groups of sizes. A human lung may have a greater number of cysts and sizes that are beyond the range simulated in this phantom [1,2,4,5,6]. Additionally, in human lungs, cysts are circular or oval shaped [3,4,6]. However, the cystic lung phantom design contains cylindrical shaped cavities due to its greater precision during the engineering phase. Despite these limitations, the lung compartments and simulated cysts were correctly identified by the CT scan and analysis pipeline, which provided valid measurements.

Overall, our cystic lung phantom serves an unmet need in the verification of CT-based cyst scoring procedures. Moreover, our model may be used to verify or calibrate post-processing corrections of the cyst measurements in the long-term, particularly amid changes in scanner platforms, imaging technologies, and operator variability.

**Author Contributions:** Conceptualization, H.W.; methodology, H.W., D.M. and E.B.; software, H.W.; validation, S.S. and F.H.; formal analysis, H.W. and S.S.; investigation, H.W.; resources, H.W. and D.M.; data curation, H.W. and S.S.; writing—original draft preparation, S.S.; writing—review and editing, H.W., S.S, and F.H.; visualization, S.S. and H.W.; supervision, H.W.; project administration, H.W and D.M.; funding acquisition, H.W. All authors have read and agreed to the published version of the manuscript.

**Funding:** This research was funded by the Division of Intramural Research, National Heart, Lung and Blood Institute, NIH Internal Research Program, the National Institutes of Health, USA, grant number HL006141.

**Conflicts of Interest:** The authors declare no conflicts of interest. The funders had no role in the design of the study; in the collection, analyses, or interpretation of data; in the writing of the manuscript; or in the decision to publish the results.

## Abbreviations

The following abbreviations are used in this manuscript:

| | |
|---|---|
| LAM | Lymphangioleiomyomatosis |
| NIST | National Institutes of Standards and Technology |
| TSC | Tuberous Sclerosis Complex |
| CT | Computed Tomography |
| mm | Millimeters |
| kVp | Kilovoltage peak |
| mA | Milliampere-seconds |
| mGy*cm | Milligray-centimeter |
| CTDI vol index | Computed tomography volume dose index |